\documentclass[%
aps,
pra,
 amsmath,amssymb,
 superscriptaddress,
preprint,%
longbibliography
]{revtex4-1}

\usepackage{graphicx}
\usepackage{dcolumn}
\usepackage{bm}
\usepackage{ulem}
\usepackage[utf8]{inputenc}
\usepackage[T1]{fontenc}
\usepackage{mathptmx}
\usepackage{textcomp} 
\usepackage{nicefrac}
\usepackage{xcolor}
\usepackage{gensymb}
\usepackage{multibib}
\usepackage{svg}

\usepackage{lineno}

\newcommand{\ve}[1]{\boldsymbol{#1}}

\usepackage{siunitx}

\begin{document}

\title[Observation of Floquet states in graphene]{Observation of Floquet states in graphene}

\author{Marco Merboldt$^\S$} %
\address{I. Physikalisches Institut, Georg-August-Universit\"at G\"ottingen, Friedrich-Hund-Platz 1, 37077 G\"ottingen, Germany}
\thanks{These authors contributed equally.}

\author{Michael Schüler$^\S$} %
\address{Laboratory for Materials Simulations, Paul Scherrer Institute, CH-5232 Villigen PSI, Switzerland}
\address{Department of Physics, 
University of Fribourg, CH-1700 Fribourg, Switzerland}
\thanks{These authors contributed equally.}

\author{David Schmitt} %
\address{I. Physikalisches Institut, Georg-August-Universit\"at G\"ottingen, Friedrich-Hund-Platz 1, 37077 G\"ottingen, Germany}

\author{Jan Philipp Bange} %
\address{I. Physikalisches Institut, Georg-August-Universit\"at G\"ottingen, Friedrich-Hund-Platz 1, 37077 G\"ottingen, Germany}

\author{Wiebke Bennecke} %
\address{I. Physikalisches Institut, Georg-August-Universit\"at G\"ottingen, Friedrich-Hund-Platz 1, 37077 G\"ottingen, Germany}

\author{Karun Gadge}%
\affiliation{Institut für Theoretische Physik, Georg-August-Universit\"at G\"ottingen, Friedrich-Hund-Platz 1, 37077 G\"ottingen, Germany}

\author{Klaus Pierz} %
\address{Physikalisch-Technische Bundesanstalt, Bundesallee 100, 38116 Braunschweig, Germany}

\author{Hans Werner Schumacher} %
\address{Physikalisch-Technische Bundesanstalt, Bundesallee 100, 38116 Braunschweig, Germany}

\author{Davood Momeni} %
\address{Physikalisch-Technische Bundesanstalt, Bundesallee 100, 38116 Braunschweig, Germany}

\author{Daniel Steil} %
\address{I. Physikalisches Institut, Georg-August-Universit\"at G\"ottingen, Friedrich-Hund-Platz 1, 37077 G\"ottingen, Germany}

\author{Salvatore~R.~Manmana}%
\affiliation{Institut für Theoretische Physik, Georg-August-Universit\"at G\"ottingen, Friedrich-Hund-Platz 1, 37077 G\"ottingen, Germany}

\author{Michael A. Sentef}
\email{sentef@uni-bremen.de}%
\address{Institute for Theoretical Physics and Bremen Center for Computational
Materials Science, University of Bremen, 28359 Bremen, Germany}
\affiliation{Max Planck Institute for the Structure and Dynamics of Matter,
Center for Free-Electron Laser Science (CFEL),
Luruper Chaussee 149, 22761 Hamburg, Germany}

\author{Marcel Reutzel} \email{marcel.reutzel@phys.uni-goettingen.de}%
\address{I. Physikalisches Institut, Georg-August-Universit\"at G\"ottingen, Friedrich-Hund-Platz 1, 37077 G\"ottingen, Germany}

\author{Stefan Mathias} \email{smathias@uni-goettingen.de}%
\address{I. Physikalisches Institut, Georg-August-Universit\"at G\"ottingen, Friedrich-Hund-Platz 1, 37077 G\"ottingen, Germany}
\address{International Center for Advanced Studies of Energy Conversion (ICASEC), University of Göttingen, Göttingen, Germany}

\begin{abstract}


Recent advances in the field of condensed-matter physics have unlocked the potential to realize and control emergent material phases that do not exist in thermal equilibrium~\cite{Basov17natmat,Torre21rmp,Bao22natrevphy}. One of the most promising concepts in this regard is Floquet engineering, the coherent dressing of matter via time-periodic perturbations~\cite{Rudner20nrp,oka_floquet_2019}. However, the broad applicability of Floquet engineering to quantum materials is still unclear. For the paradigmatic case of monolayer graphene, the theoretically predicted Floquet-induced effects~\cite{Oka09prb,Usaj14prb,Sentef15natcom,Schueler20prx}, despite a seminal report of the light-induced anomalous Hall effect~\cite{McIver20natphys}, have been put into question~\cite{Aeschlimann21nanoletters}. Here, we overcome this problem by using electronic structure measurements to provide direct experimental evidence of Floquet engineering in graphene. We report light-matter-dressed Dirac bands by measuring the contribution of Floquet sidebands, Volkov sidebands~\cite{baggesen2008theory,saathoff_laser-assisted_2008}, and their quantum path interference~\cite{mahmood_selective_2016,park_interference_2014} to graphene's photoemission spectral function. Our results finally demonstrate that Floquet engineering in graphene is possible, paving the way for the experimental realization of the many theoretical proposals on Floquet-engineered band structures and topological phases~\cite{Hubener17natcom,Lindner11natphys,Claassen16natcom,Zhang16prb,Yan16prl,Sentef15natcom,Liu18prl,Rodriguez21ap,Topp19prr,Katz20prb,Kenness21natphys,Chan23pnas}.

\end{abstract}

\def\thefootnote{*}\footnotetext{These authors contributed equally to this work}\def\thefootnote{\arabic{footnote}}
\maketitle


\noindent\textbf{Introduction}

The field of topological Floquet engineering was started by Oka and Aoki~\cite{Oka09prb}, who proposed that the Haldane model~\cite{Haldane88prl} -- one of the most paradigmatic models of topology in condensed matter physics -- can be realized in monolayer graphene. Upon irradiation with circularly polarized light, a dynamical topological anomalous Hall state can be induced, which manifests itself via the formation of Floquet replicas of the original Bloch bands and a band gap opening at the Brillouin zone's K and K$^\prime$ points with an inherent change of the Chern number. Such a light-induced topological phase transition can be rationalized within Floquet theory~\cite{Shirley65prb,Usaj14prb,Sentef15natcom}, and has been demonstrated experimentally for ultracold fermions in an optical lattice~\cite{Jotzu14nat} and photonic crystals~\cite{Rechtsman13nat}. Specifically in graphene, the power and versatility of Floquet engineering was demonstrated in a pioneering experiment reporting a light-induced anomalous Hall effect~\cite{McIver20natphys}. However, significant theoretical effort was needed to explain the experimental findings by a combination of Floquet-induced topology and asymmetric photocarrier distributions~\cite{Sato19prb,Nuske20prr}. Moreover, while there is a growing body of experimental reports highlighting the applicability of Floquet engineering to condensed matter systems~\cite{Wang13sci,sie14natmat,mahmood_selective_2016,Reutzel20natcom,Aeschlimann21nanoletters,Park22nat,Zhou23nat,Ito23nat,Kobayashi23natphys,Zhou23prl}, for graphene, today's preconception is that the ultrafast decoherence time of only a few tens of femtoseconds~\cite{Heide21nanoletters} is so short that it hinders the efficient generation and verification of Floquet-engineered phases~\cite{Aeschlimann21nanoletters,Sato20prb,Broers22prr}. Hence, any unambiguous experimental demonstration of the Floquet engineering concept for monolayer graphene would be a major advancement in the field.

In this article, we use linearly-polarized infrared driving light fields to coherently dress monolayer graphene and then probe the energy-momentum dispersion of the light-dressed band structure with extreme ultraviolet laser pulses generated via high-harmonic generation in a new type of angle-resolved photoemission spectroscopy experiment (ARPES), known as momentum microscopy~\cite{medjanik_direct_2017,Keunecke20timeresolved}. In the direct comparison of measured and calculated ARPES maps of the light-dressed band structure, we identify energy- and momentum-resolved fingerprints of Floquet sidebands, Volkov sidebands, and their mutual interference. Specifically, we show that the latter, i.e., the quantum path interference between Floquet and Volkov states in momentum space, is a powerful tool to unambiguously identify light-dressed band structures with time-resolved momentum microscopy. Our work opens up a direct pathway to test the many theoretical proposals of light-induced  phase transitions that have remained elusive experimentally to this date~\cite{Hubener17natcom,Lindner11natphys,Claassen16natcom,Zhang16prb,Yan16prl,Sentef15natcom,Liu18prl,Rodriguez21ap,Topp19prr,Katz20prb,Kenness21natphys,Chan23pnas}.


\vspace{1cm}
\noindent\textbf{Experimental observables for Floquet engineering in graphene}

\begin{figure}[b]
    \centering
     \includegraphics[width=.95\linewidth]{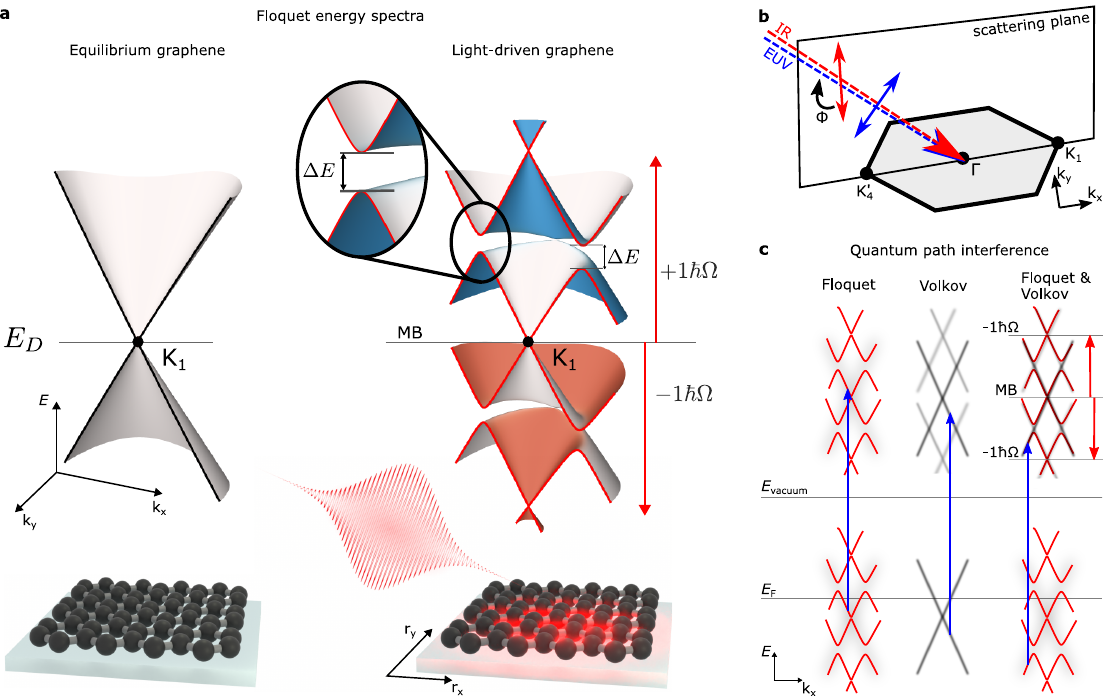}
    \caption{\textbf{Floquet-engineered modifications to graphene's electronic structure that are accessible in ARPES experiments.}
    \textbf{a} Energy- and in-plane momentum-resolved representation of the energy eigenstates of graphene in equilibrium (black) and light-driven graphene (red) in the proximity of the K$_1$ point. The characteristics of the Floquet-engineered band structure are (i) the formation of sidebands $\pm1\hbar\Omega$ spaced by the driving photon energy from the original main band (MB), and (ii) the formation of energy gaps $\Delta E$ where sidebands of different photon order cross.
    \textbf{b} The plane-of-incidence (scattering plane) of the laser fields is along the K$^\prime_4$-$\Gamma$-K$_1$ crystal direction. The EUV probe pulses are $p$-polarized and the linear polarization of the IR driving laser pulses is tuneable via the polarization angle $\Phi$.    
    \textbf{c} The coherent interaction of the IR laser pulses within graphene and with the quasi-free photoelectrons after the photoemission process lead to the formation of Floquet (red) and Volkov (grey) sidebands, respectively. Quantum path interference between Floquet and Volkov transitions have to be considered, as both states are probed at the same photoelectron energy and in-plane momentum.    
    }
\end{figure}

Figure~1a shows a schematic of the equilibrium (black) and the Floquet-engineered (red) electronic structure of graphene (using linearly-polarized laser pulses). As discussed in many earlier reports~\cite{Oka09prb,Usaj14prb,Sentef15natcom,Schueler20prx}, the light-dressed band structure deviates from its equilibrium counterpart based on two distinct signatures: First (i), Floquet theory predicts that higher-order photon-dressed sidebands of the main band Dirac cone are formed (labelled $\pm1\hbar\Omega$ and MB in Fig.~1a, respectively). Second (ii), energy bands are gapped where sidebands of different photon order cross and hybridize in energy-momentum space ($\Delta E$ in Fig.~1a). To probe such coherent modifications of graphene's band structure, the combination of ARPES, in particular momentum microscopy, with a femtosecond pump-probe setup is ideally suited~\cite{Wang13sci,mahmood_selective_2016,Reutzel20natcom,Aeschlimann21nanoletters,Zhou23nat,Ito23nat,Zhou23prl,Keunecke20prb}. While an infrared (IR) laser pulse enables the periodic driving of the system, a time-delayed extreme ultraviolet (EUV) laser pulse allows to record the energy- and in-plane momentum-resolved photoemission spectral function. Notably, as broadband ultrashort laser pulses are used in time-resolved ARPES experiments, Floquet energy gaps $\Delta E$ might not be directly resolvable (extended Fig.~\ref{fig:floquetenergygaps}). Hence, it is more straightforward to study the aforementioned case (i), i.e., the photon-dressed sideband formation in momentum space. We opt for this route by using our in-house photoemission endstation~\cite{Keunecke20timeresolved,Schmitt22nat} that combines a time-of-flight momentum microscope~\cite{medjanik_direct_2017} with an ultrafast table-top EUV light source. From the $n$-doped graphene sample grown on 4H-SiC~\cite{momeni2018minimum,Duvel22nanolett}, the momentum microscope facilitates the collection of photoelectrons as a function of energy $E$ and both in-plane momenta k$_x$ and k$_y$. In our experimental geometry, the EUV probe (26.5\,eV, $20$\,fs, $p$-polarized) and infrared (IR) driving ($\hbar\Omega = 0.65$\,eV, 100\,fs, 3~MV/cm) pulses impinge nearly co-linearly onto the graphene sample at an oblique angle of incidence of 22°. The K$^\prime_4$-$\Gamma$-K$_1$ crystal direction lies in the scattering plane and photoelectrons are collected in the proximity of the K$_1$ point (Fig.~1b).


\vspace{.5cm}
\noindent\textbf{IR polarization dependence of sideband photoemission yield}

\begin{figure}
    \centering
    \includegraphics{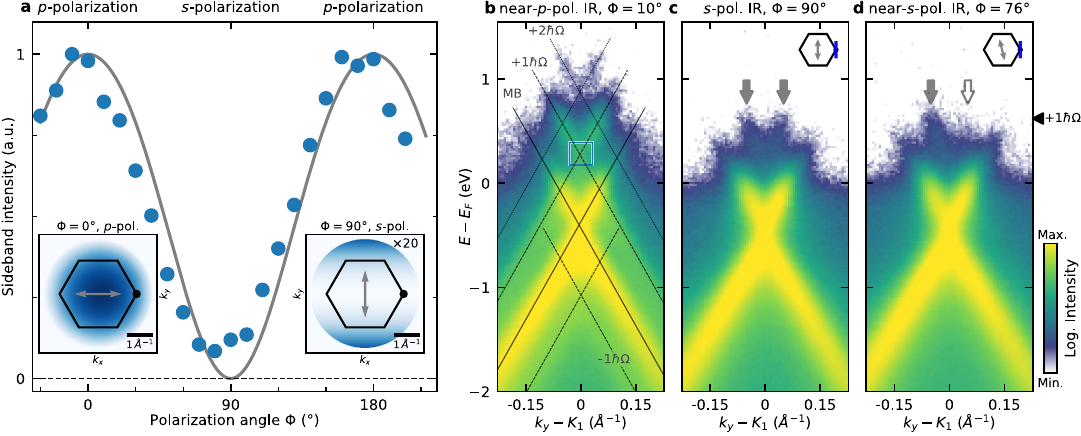}
    \caption{\textbf{IR polarization dependence of the sideband photoemission spectral weight.}
    \textbf{a} IR polarization dependent $+1\hbar\Omega$ sideband photoemission yield integrated over the blue region of interest in \textbf{b}. The insets show the momentum-resolved Volkov sideband amplitude calculated within the Volkov formalism for $p$- (left) and $s$-polarized (right) IR pulses [blue-shaded color code; white corresponds to no sideband amplitude; calculated with equation~(\ref{eq:alpha})]. Note that the color scale for $s$-polarized IR pulses is multiplied by a factor 20. The corners of the hexagons indicate the momenta of the six K (K$^\prime$) points of graphene; the K$_1$ point is marked with a black circle. The doubleheaded grey arrows indicate the direction of the in-plane IR electric field component. From this momentum-resolved analysis of the Volkov formalism, the grey line in the main panel is calculated and shows the Volkov sideband amplitude for the Dirac cone at the K$_1$ point for all polarizations $\Phi$.
    \textbf{b,c,d} Energy- and momentum-resolved photoemission data collected in temporal overlap of the EUV and the IR pulses for $k_x\approx0$~\AA$^{-1}$ and $\Phi=10$° (near-$p$-polarized), 90° ($s$-polarized), and 76° (near-$s$-polarized). The vertical grey arrows in \textbf{d} highlight the $\pm k_y$ asymmetry of the $+1\hbar\Omega$ sideband spectral weight for near-$s$-polarized IR pulses. The black arrowhead in \textbf{d} denotes the center energy of the momentum maps shown in Fig.~3.
    } 
\end{figure}

During the presence of the IR driving laser field, Floquet eigenstates are created and can, in principle, be photoexcited by the EUV laser pulse and measured with ARPES (Fig.~1c, left panel). However, when ARPES is used for such measurements, it is well-known that a competing process can lead to similar photoemission signatures: The coherent interaction of quasi-free photoelectrons with the driving light field leads to the formation of so-called Volkov sidebands of the main photoemission signature (Fig.~1c, middle panel)~\cite{baggesen2008theory,saathoff_laser-assisted_2008}. Importantly, these Volkov sidebands look similar to Floquet sidebands in ARPES, but are \textit{not} an indication for a light-dressed band structure. However, because the Floquet and the Volkov transitions are observed at the same final state energy, quantum path interference effects between both excitation pathways occur and can be observed ~\cite{park_interference_2014,mahmood_selective_2016} (Fig.~1c, right panel). Hence, for the unambiguous identification of Floquet sidebands in ARPES, the experimental challenge lies in the discrimination of Floquet sidebands, Volkov sidebands, and their interference pattern.

For this task, we make use of the fact that the (k$_x$,k$_y$)-momentum-resolved amplitude of Volkov sidebands can be controlled by varying the polarization $\Phi$ of the IR field~\cite{Keunecke20prb,park_interference_2014,mahmood_selective_2016}, and do not yet consider possible contributions of Floquet states (methods): For $p$-polarized IR pulses ($\Phi=0$°), the surface projected electric field vector $\ve{E}$ is oriented parallel to the scattering plane, and the amplitude of Volkov sidebands is large for all six K$_{i}$ points (Fig.~2a, left inset). In contrast, for $s$-polarized IR pulses ($\Phi=90$°), the surface projected electric field vector $\ve{E}$ is oriented perpendicular to the scattering plane, and the Volkov sideband amplitude vanishes for all momenta along the K$^\prime_4$-$\Gamma$-K$_1$ crystal direction (Fig.~2a, right inset). Based on these calculations, in the main panel of Fig.~2a, we plot the calculated polarization dependence of the Volkov sideband amplitude at the momentum of the K$_1$ point (grey line): The Volkov sideband amplitude is maximized for $p$-polarized light and vanishes for $s$-polarized light. Hence, if it is possible to observe sideband photoemission spectral weight in the proximity of the K$_1$ point for $s$-polarized IR light, where the Volkov sideband amplitude must be zero, the result would be indicative for the formation of Floquet states.

Figures 2b,c,d show energy- and momentum-resolved photoemission data collected in temporal overlap of the EUV and the IR laser pulses; the polarization of the IR pulses is $\Phi=10$° (near-$p$-polarization), 90° ($s$-polarization), and 76° (near-$s$-polarization). Starting with the case of near-$p$-polarization, we can clearly identify the main Dirac cone (main band, MB), and, in addition, sidebands spaced by $-1\hbar\Omega$, $+1\hbar\Omega$, and $+2\hbar\Omega$ (Fig.~2b). In order to evaluate the impact of the IR polarization on the sideband photoemission yield, we systematically evaluate the data in a 0.064~\AA$^{-1}\times0.064$~\AA$^{-1}\times0.22$~eV region of interest at the $+1\hbar\Omega$ sideband (blue box in Fig.~2b). For all polarizations $\Phi$, we detect a spectral weight originating from the $+1\hbar\Omega$ sideband (Fig.~2a, data points). Notably, even for the case of $s$-polarized IR pulses, we still identify spectral weight of the $+1\hbar\Omega$ sideband (Fig.~2c). This observation is in stark contrast to the expected fingerprints in the hypothetical case of Volkov states only, which should not lead to a finite $+1\hbar\Omega$ sideband photoemission yield for $s$-polarized IR pulses (grey line in Fig.~2a).

The natural follow-up question is whether the $+1\hbar\Omega$ sideband photoemission signal collected for $s$-polarized IR light already constitutes unambiguous proof for the existence of Floquet states (Fig.~2c), especially since such a signature could not be identified in a previous ARPES experiment~\cite{Aeschlimann21nanoletters}. For instance, it might be possible that the linear polarization of the IR pulses is not pure, which would also lead to a finite Volkov sideband intensity. Hence, an additional hallmark that supports the successful generation of Floquet states seems necessary. Looking again at Fig.~2b-d, we find another very strong signature in the energy-momentum-resolved photoemission spectra that is incompatible with the Volkov picture alone: While the near-$p$- and the $s$-polarized measurements show symmetric $+1\hbar\Omega$ sideband spectral weight in $\pm k_y$ momentum-direction (Fig.~2b,c), for $\Phi=76$°, the measurement exhibits a striking $\pm k_y$ asymmetry between the two sides of the Dirac cone (Fig.~2d, vertical arrows). In the following, we will show that such strongly asymmetric photoemission signatures of the observed sidebands cannot be explained within the Volkov or the Floquet picture alone, but must be a result of quantum path interference of Volkov and Floquet states.


\vspace{.5cm}

\noindent\textbf{Verification of Floquet states in graphene}

\begin{figure}
    \centering
    \includegraphics[width=0.9\linewidth]{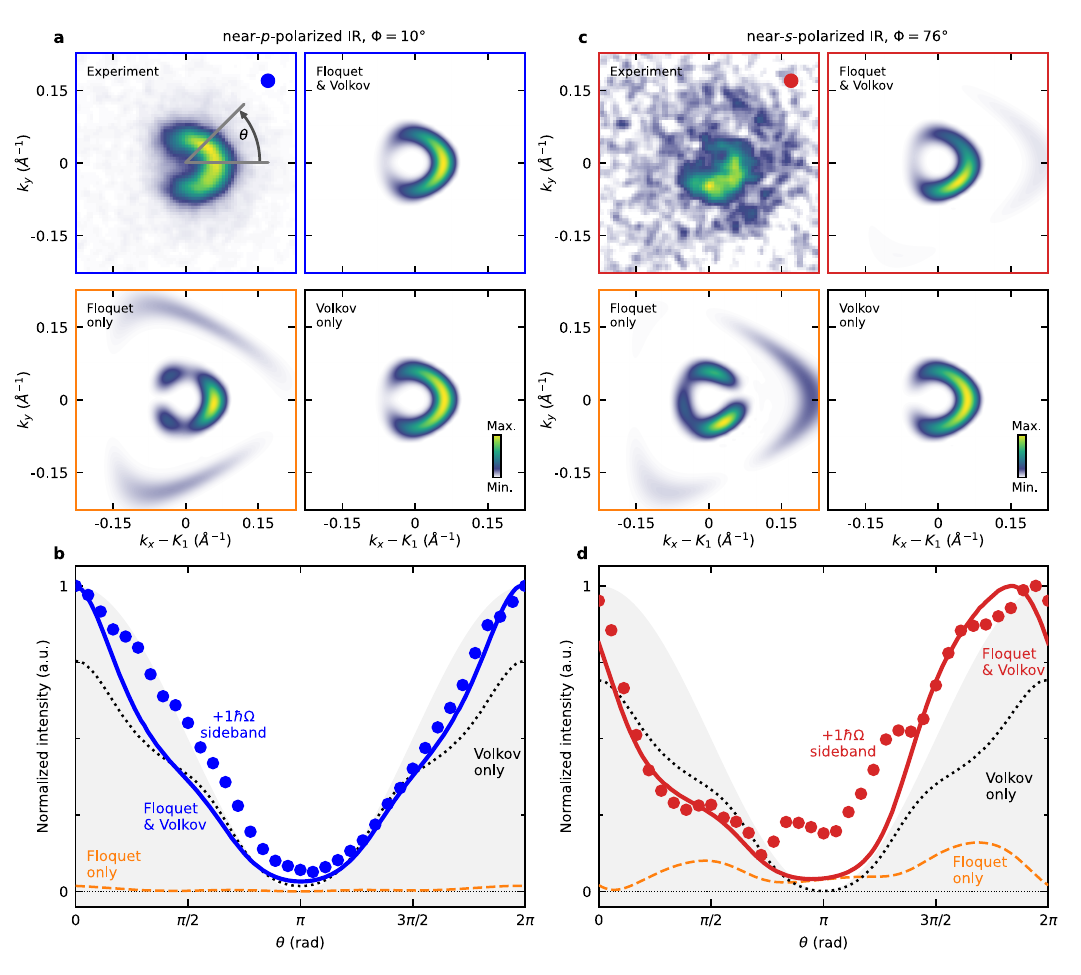} 
    \caption{\footnotesize\textbf{Sideband photoemission momentum-fingerprints and quantum path interference between Floquet and Volkov transitions.}
    \textbf{a,c} Measured and calculated $+1\hbar\Omega$ sideband momentum-fingerprints for near-$p$-polarized ($\Phi=10$°, \textbf{a}) and near-$s$-polarized ($\Phi=76$°, \textbf{c}) IR pulses (E-E$_{\rm F}=0.63$~eV, $\Delta t=0$~fs, $+1\hbar\Omega$ labelled arrowhead in Fig.~2d). The theory maps contain the coherent sum of Floquet and Volkov transitions (blue [red] axes in (\textbf{a}) [(\textbf{c})]), Floquet transitions only (orange axes), and Volkov transitions only (black axes).
    \textbf{b,d} The momentum-fingerprint of the $+1\hbar\Omega$ sideband spectral weight is parameterized with the angle $\Theta$ [cf. panel (\textbf{a})]. 
    \textbf{b} For near-$p$-polarized IR pulses, the measured photoemission momentum-fingerprint of the $+1\hbar\Omega$ sideband (blue circles) qualitatively mirrors the $\Theta$-dependence of the dark corridor (grey area), and is well-described by the calculations that consider Floquet and Volkov transitions (blue line).
    \textbf{d} In the case of near-$s$-polarized IR pulses, the measured photoemission momentum-fingerprint of the $+1\hbar\Omega$ sideband (red circles) is qualitatively different to the $\Theta$-dependence expected from the dark corridor, and shows a distinct asymmetry with a minimum and a maximum for $\Theta<\pi$ and $\Theta>\pi$, respectively. Neither the Volkov-only (black dotted line) nor the Floquet-only (orange dashed) calculations predict this strong asymmetry. Agreement between experiment and theory can only be found if quantum path interference processes between Floquet and Volkov transitions are considered (red line).
    }
\end{figure}

To investigate the $\pm k_y$ momentum asymmetry in more detail, we make use of the full momentum-resolved data collection capability of our photoemission endstation and generate ($k_x$,$k_y$)-resolved photoemission maps at the energy of the $+1\hbar\Omega$ sideband (Fig.~3a,c). While the momentum map of the near-$p$-polarized case ($\Phi=10$°) shows the well-known horseshoe-like spectral weight distribution in the sideband that originates from the dark corridor effect~\cite{Shirley95prb} (Fig.~3a), in the case of near-$s$-polarized IR pulses ($\Phi=76$°), the spectral weight distribution is dominated by the strong $\pm k_y$ asymmetry (Fig.~3b) that was already observed in the energy-momentum-resolved data in Fig.~2d. In order to show that this asymmetry provides unambiguous evidence for Floquet states in graphene, we compare our experimental data with ARPES momentum maps calculated within the time-dependent non-equilibrium Green's function formalism (methods). The calculations are performed such that the spectral weight in the momentum maps can contain contributions of the coherent sum of Floquet and Volkov sidebands, Floquet sidebands only, or Volkov sidebands only, as labelled in the respective momentum maps in Fig.~3a,c. Intriguingly, already from the visual inspection of the momentum maps, it is obvious that the experimental $\pm k_y$ momentum asymmetry for the near-$s$-polarized IR pulses can only be reproduced by the calculations that consider constructive and destructive quantum path interference processes between Floquet and Volkov transitions (Fig.~3c, red highlighted momentum maps). Neither the calculated momentum-dependence of the pure Floquet (orange) nor of the pure Volkov (black) transitions can reproduce the experimentally observed asymmetry.

We evaluate the momentum-dependent spectral weight distribution of the $+1\hbar\Omega$ sideband in experiment and theory by integrating the sideband signal in segments around the K$_1$ point  parameterized with the angle $\Theta$ (cf. Fig.~3a). For near-$p$-polarized IR pulses (Fig.~3b), we make two important observations: First, we find that the experimental data (blue circles) follows the $\Theta$-dependence expected from the dark corridor effect~\cite{Shirley95prb} (grey area), indicating that the $+1\hbar\Omega$ sideband spectral weight is a near-perfect replica of the non-driven main band (extended Fig.~\ref{fig:crosscorrelation}c). Second, the experimental $\Theta$-dependence is well-described by the calculations that include the coherent contribution of Floquet and Volkov states (blue solid line). However, by comparing the $\Theta$-dependent spectral weight of the Volkov only (black dotted line) and the Floquet only calculation (orange dashed line), it is obvious that mainly Volkov transitions contribute to the measured $+1\hbar\Omega$ sideband intensity, as expected for near-$p$-polarized driving.

Next, we repeat the same evaluation for the $+1\hbar\Omega$ sideband momentum pattern for the case of near-$s$-polarized IR pulses ($\Phi=76$°, Fig.~3c,d). In Fig.~3d, it is directly clear that the sideband's photoemission spectral weight (red circles) deviates from the pure cosine-like $\Theta$-dependence and thus does not follow the periodicity of graphene's dark corridor (grey area). To verify the contribution of Floquet states at $\Phi=76$° IR driving, Fig.~3d shows the $\Theta$-dependence of the spectral weight of the calculated momentum maps for Floquet-only (orange line), Volkov-only (black line), and the case of photoemission quantum path interference of Floquet and Volkov states (red line). Notably, the $\Theta$-dependence of the Volkov-only solution is close to symmetric for $\Theta\lessgtr\pi$, as found for the case of near-$p$-polarized IR driving, but in contrast to the experimental data of the $\Phi=76$° case. Hence, the measured momentum-asymmetric $+1\hbar\Omega$ sideband spectral weight distribution cannot be described within the Volkov formalism alone. Likewise, the calculated Floquet-only momentum-dependence is close to symmetric for $\Theta\lessgtr\pi$, and therefore does not reproduce our experimental observations either. However, in the case that constructive and destructive quantum path interference processes of Floquet and Volkov transitions are considered, our calculations clearly reproduce the experimentally observed asymmetry (red line). In other words, the strong asymmetric momentum-fingerprint of the $+1\hbar\Omega$ sideband intensity can only be observed if Floquet and Volkov states are detected, which thus, for the first time, directly verifies the experimental realization of Floquet engineering in graphene.

\vspace{.5cm}
\noindent\textbf{Quantum path interference of Floquet and Volkov states}

Our results indicate that the momentum-resolved sideband photoemission spectral weight is dependent on the relative phase of the Floquet and Volkov transitions contributing to the quantum path interference conditions. As discussed by Park~\cite{park_interference_2014}, the phase of the Floquet amplitude is determined by the projection of the IR field onto the momentum $\mathbf{k}$, while the Volkov phase exhibits a much weaker dependence. Therefore, as in an interferometer where the phase of one channel can be controlled, it must be possible to flip the asymmetric momentum fingerprint of the $+1\hbar\Omega$ sideband by controlling the polarization angle around $\Phi$ = 90° ($s$-polarization). In Fig.~4a-c, we show measured (top row) and calculated (bottom row) momentum maps of the $+1\hbar\Omega$ sideband for $\Phi$ = 90° (Fig.~4b) and $\Phi$ = 90° $\mp$ 4° (Fig.~4a,c). In the case of $s$-polarized IR pulses, the surface projected electric field vector is oriented perpendicular to the scattering plane and the quantum path interference conditions are symmetric in $+k_y$ and $-k_y$ direction (Fig.~4b). In contrast, if $\Phi\neq 90$°, the surface projected electric field vector and the scattering plane are not perpendicular anymore with respect to each other, and quantum path interference leads to an asymmetric spectral weight for $\pm k_y$ (Fig.~4a,c). In particular, the asymmetry flips for angles $\lessgtr90$°, as expected from theory. Finally, Fig.~4d shows the systematic evaluation of the momentum asymmetry $A=\left(I_{\rm +k_y}-I_{\rm -k_y}\right)/\left(I_{\rm +k_y}+I_{\rm -k_y}\right)$ as a function of the IR pulse polarization angle $\Phi$. In agreement between experiment (blue dots) and theory (black line), we find that the asymmetry $A$ increases from the $p$-polarized ($\Phi=0$°, 180°) to the $s$-polarized case ($\Phi=90$°). Close to $s$-polarization ($\Phi=90$°), the asymmetry $A$ flips and is most sensitive to changes in polarization, as both Floquet and Volkov sideband contributions have comparable amplitudes.

\begin{figure}
    \centering
    \includegraphics[width=\linewidth]{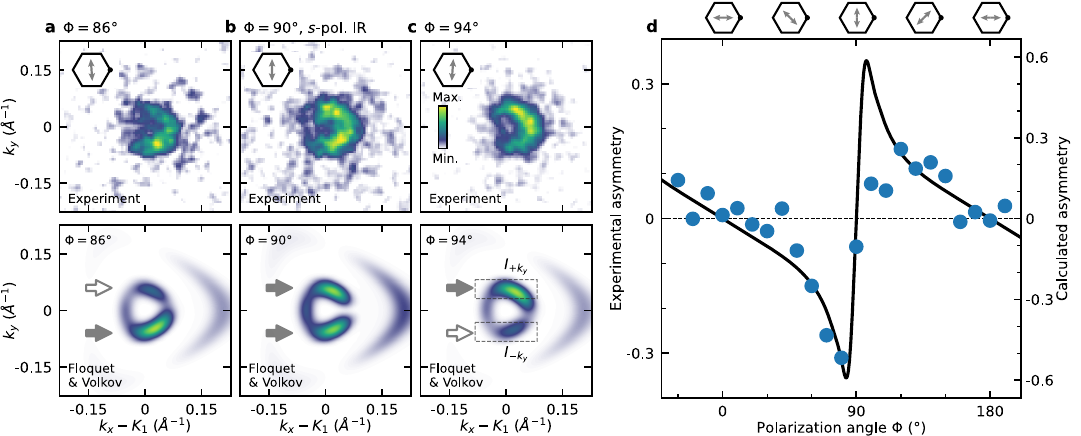} 
    \caption{\textbf{Control of quantum path interference between Floquet and Volkov states via the IR pulse polarization $\Phi$.}
    \textbf{a-c} The upper and the lower row show measured and calculated photoemission momentum maps of the $+1\hbar\Omega$ sideband, respectively (E-E$_{\rm F}=0.63$~eV, $\Delta t=0$~fs). As the polarization of the IR pulses (doubleheaded arrows in inset) is rotated from $\Phi=90\degree-4\degree=86$° (\textbf{a}) to $\Phi=90$° (\textbf{b}) and $\Phi=90\degree+4\degree=94$° (\textbf{c}), the spectral weight is more intense for $-k_y$, similarly intense in $\pm k_y$ and more intense for $+k_y$, respectively (cf. horizontal arrows in lower panels).
    \textbf{d} Systematic evaluation of the $\Phi$-dependent asymmetry $A=\left(I_{\rm +k_y}-I_{\rm -k_y}\right)/\left(I_{\rm +k_y}+I_{\rm -k_y}\right)$ (regions of interest indicated in (\textbf{c})). The data points and the black line are extracted from experiment and theory, respectively. The asymmetry $A$ changes sign when the surface projected electric field vector (double-headed arrow) is rotated through $\Phi=0$°, 90°, and 180°.
    } 
\end{figure}


\vspace{.5cm}

\noindent\textbf{Conclusions and outlook}

We directly and unambiguously demonstrate the successful generation of Floquet states in graphene. We do so by exploiting quantum path interference processes of Floquet and Volkov transitions in the time- and momentum-resolved photoemission experiment, i.e., without the need to resolve vanishingly small energy gaps or band renormalizations. More generally, our results are the first direct experimental proof for the seminal theoretical predictions of Floquet states in monolayer graphene, first proposed about 15 years ago~\cite{Oka09prb}. In a next step, the application of circularly polarized driving light pulses will show if a phase transition to a topologically non-trivial state can be achieved. Moreover, since the pioneering work of Oka and Aoki~\cite{Oka09prb}, many subsequent theoretical proposals have extended the Floquet engineering concept to quantum materials such as Weyl semimetals~\cite{Hubener17natcom} and twisted heterostructures of two-dimensional materials~\cite{Rodriguez21ap,Topp19prr,Katz20prb}, motivating even the combination of Floquet engineering and twistronics~\cite{Kenness21natphys}. Our results will allow the experimental verification of all these theoretical proposals in the coming years, promising the creation of light-matter coupled material phases without a counterpart in thermal equilibrium.

\section{ACKNOWLEDGEMENTS}

We thank G.~S.~Matthijs~Jansen for fruitful discussions. The Göttingen authors acknowledge funding by the Deutsche Forschungsgemeinschaft (DFG, German Research Foundation) via 217133147/SFB 1073, projects B03, B07, and B10. The PTB group was supported by the Deutsche Forschungsgemeinschaft (DFG) project Pi385/3-1 within the Research Unit FOR5242 and the DFG Germany’s Excellence Strategy–EXC-2123 QuantumFrontiers – 390837967. M.S. acknowledges support from SNSF Ambizione Grant No. PZ00P2-193527. M.A.S. was funded by the European Union (ERC, CAVMAT, project no. 101124492). We acknowledge access to Piz Daint at the Swiss National Supercomputing Centre, Switzerland under the PSI's share with the project ID psi10.

\section{AUTHOR CONTRIBUTIONS}
M.M., D.Sch., J.P.B., and W.B. carried out the time-resolved momentum microscopy experiments. M.M. analyzed the experimental data with contributions from D.Sch.. M.S. performed the calculations. K.P., H.W.S., and D.M. provided the graphene sample. All authors discussed the results. S.M. and M.R. were responsible for the overall project direction and wrote the manuscript with contributions from all co-authors.


\clearpage

\section*{Methods}
\clearpage

\setcounter{section}{0}

\section{Experimental setup and sample preparation}

\begin{figure}[bt]
    \centering
    \includegraphics[width=\linewidth]{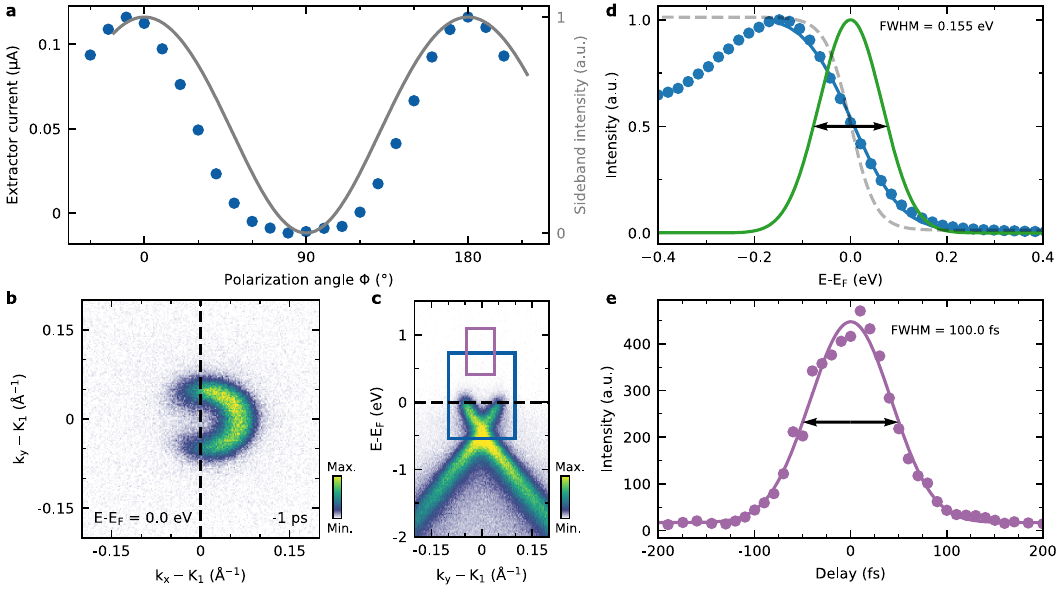}
    \caption{\footnotesize\textbf{Characterization of the IR pulse polarization $\Phi$ on the sample, ARPES data of non-driven graphene, and the time- and energy-resolution of the experiment.}
    \textbf{a} Irradiation of the graphene sample with IR laser pulses leads to a measurable current between the sample and the extractor, which varies as the pulse's polarization $\Phi$ changes from $s$ to $p$ (data points). As a guide-to-the-eye, the gray line is reproduced from Fig.~2a and shows the calculated Volkov sideband amplitude at the K$_1$ point.
    \textbf{b, c} Energy-momentum-resolved (at k$_{\rm x}$-K$_1$ = 0 \AA$^{-1}$, indicated by dashed line in b)) and momentum-momentum-resolved (E-E$_{\rm F}=0$ eV, indicated by dashed line in c)) photoemission data collected when the IR and EUV pulses are far from temporal overlap ($\Delta t=-1$~ps).
    \textbf{d} Momentum-filtered energy distribution curve (data points) in the region of interest indicated in panel \textbf{c}, blue box. The fit (blue solid line) is carried out with a Fermi-Dirac distribution at 300~K (grey dashed line) convoluted with a Gaussian with a width of 0.155~eV (FWHM, black horizontal line).
    \textbf{e} Quantification of the cross-correlation of the IR driving and the EUV probing laser pulses. The data points are obtained by filtering the photoemission yield from the $+1\hbar\Omega$ sideband in the magenta 0.7~eV $\times$ 0.07~\AA$^{-1}$ (k$_x$) $\times$ 0.07~\AA$^{-1}$ (k$_y$) region of interest (both pulses $p$-polarized). The width of the cross-correlation is 100$\pm3$~fs (FWHM, black horizontal line).}
    \label{fig:crosscorrelation}
\end{figure}

The time-resolved momentum microscopy experiments have been performed with our in-house photoemission endstation that combines a time-of-flight momentum microscope~\cite{medjanik_direct_2017} (Surface Concept GmbH) and a 300~W fiber laser system (Active Fiber Systems)~\cite{Keunecke20timeresolved}. The laser is operated at 500~kHz and drives a table-top high-harmonic generation (HHG) beamline and an optical parametric amplifier (OPA, Light Conversion), as detailed in refs.~\cite{Keunecke20timeresolved,Schmitt22nat}. The HHG beamline is operated with 5\,W, 50\,fs, 515\,nm pulses focused into Argon gas, and the 11th harmonic (26.5\,eV, $p$-polarized) is selected with a pair of EUV multilayer mirrors. The OPA is operated at 40~W input power (1030~nm, 220~fs) and generates the 210~mW (at sample), 100~fs, 0.65~eV IR driving laser pulses. The polarization angle $\Phi$ of the IR pulses is varied with an achromatic half-wave plate (B. Halle, 700-2500~nm). The polarization $\Phi$ of the IR pulses on the sample is determined by monitoring the extractor current as a function of the IR polarization angle (extended Fig.~\ref{fig:crosscorrelation}a) and the direct comparison of the sideband momentum fingerprints in experiment and theory (Fig.~4).

All experiments are performed at room temperature on a $n$-doped graphene sample grown on 4H-SiC~\cite{momeni2018minimum,Duvel22nanolett}. In ultra-high-vacuum, the graphene sample was annealed for 1~h at 450~°C. For all experiments, the microscope was aligned such that a momentum area with a diameter of ~1.3~\AA$^{-1}$ centered on the K$_1$ point is projected onto the detector~\cite{Keunecke20timeresolved}. In addition, the multi-dimensional photoemission data shown throughout the text are corrected for distortions in energy and momentum~\cite{xian2019symmetry,Bange232DMaterials,Bennecke24natcom}. The vacuum electric field strength of the IR pulses is approximated to 3~MV/cm with an estimated $1/e^2$ diameter of the IR beam of 300~$\mu\mathrm{m} \times 250$~$\mu$m and a peak fluence of 1.23~mJ/cm$^{2}$.


\section{Time- and energy-resolution of the photoemission experiment}

We quantify the energy resolution of the momentum microscopy experiment by fitting a momentum-filtered energy distribution curve with a Fermi-Dirac distribution broadened by a Gaussian distribution (extended Fig.~\ref{fig:crosscorrelation}b,c,d). The data is taken from a measurement at -1~ps delay, i.e., the IR and  EUV pulses are on the sample but not in temporal overlap (extended Fig.~\ref{fig:crosscorrelation}b,c). Keeping the temperature of the Fermi-Dirac distribution fixed to 300~K, the fit yields a Gaussian width of $155\pm47$~meV (extended Fig.~\ref{fig:crosscorrelation}d). The Gaussian width then describes the energy resolution of our experiment and contains contributions from the spectral width of the laser pulses and the energy resolution of the momentum microscope~\cite{Keunecke20timeresolved}. 

In extended Fig.~\ref{fig:crosscorrelation}e, we evaluate the pump-probe delay dependence of the $+1\hbar\Omega$ sideband photoemission signal in order to extract the cross-correlation of the IR and EUV beams to 100$\pm3$~fs (FWHM). With a pulse duration of the EUV pulses of 20$\pm5$~fs~\cite{Keunecke20timeresolved,Bange24SciAdv}, we extract the pulse duration of the IR pulse to 98$\pm4$~fs. This translates to a Fourier-limited spectral width of 27$\pm1$~meV and 124$\pm30$~meV for the IR and the EUV laser pulses, respectively.

\clearpage


\section{Momentum- and polarization-dependent Volkov sideband amplitude}

In the following, we briefly describe the polarization- and momentum-dependence of the Volkov sideband yield, as discussed in the main text and shown in Fig.~2a (insets and grey line). Details on this analysis can be found in refs.~\cite{park_interference_2014,madsen2005strong,baggesen2008theory,Keunecke20prb}, and we follow the earlier work of some of us (Keunecke \textit{et al.}~\cite{Keunecke20prb}). The photoelectron momentum-distribution of the first order Volkov sideband is given by 
\begin{equation}
I_1\left(k_{xy}, \theta_k,k_{z}\right)\sim I'_0\left(k_{xy}, \theta_k,k_{z}\right)\times|\alpha_{1}|^2,
\label{eq:yield}
\end{equation} 
with $I'_0\left(k_{xy},\theta_k ,k_{z}\right)$ being the photoemission yield of the undriven system, and $|\alpha_1|^2$ is the Volkov sideband amplitude. In the electron scattering description~\cite{park_interference_2014,madsen2005strong,baggesen2008theory}, the $\alpha$ parameter can be expressed as
\begin{equation}
\alpha \sim \left(\frac{e}{m_e\Omega^2}(E_{xy}k_{xy}\cos(\theta_{k}-\theta_{E})+E_{z}k_{z})\right).
\label{eq:alpha}
\end{equation}
Here, the in-plane electric field components and the in-plane momentum components are expressed in polar coordinates, i.e. $\theta_k = \tan^{-1} (k_y/k_x)$ (measured from the $\Gamma$ point) and $\theta_E = \tan^{-1}(E_y/E_x)$, as detailed in ref.~\cite{Keunecke20prb}. Moreover, $e$, $m_e$, and $\Omega$ are the elementary charge, the elementary mass, and the driving light frequency, respectively. In the insets of Fig.~2a, the momentum-dependent distribution of $\left|\alpha_1\right|^2$, which describes the Volkov amplitude, is plotted for $s$- and $p$-polarized IR pulses in our experimental geometry (see Fig.~1b and ref.~\cite{Keunecke20prb}).

\clearpage


\section{Details on the calculations}

\subsection{Experimental parameters that enter the calculations}

For the determination of the experimental parameters that enter our simulations, we start with the measured vacuum field strength of $E_0\approx 3$~MV/cm (see section~I of the methods). Further, we need to adjust the electric field strength at the surface and inside the graphene, because the interface between the graphene and the vacuum is not a sharp interface as assumed for the Fresnel equations. To do so, we introduce scaling factors $f_V$ and $f_F$ for the Volkov and Floquet field strengths at the surface and inside the graphene, respectively.

The electric field strength that generates Volkov sidebands at the surface is then
\begin{align}
    \mathbf{E}_\mathrm{eff}(\Phi) = f_V \left(\mathbf{E}_\mathrm{in}(\Phi) + \mathbf{E}_\mathrm{r}(\Phi) \right) \ ,
\end{align}
where $\mathbf{E}_\mathrm{in}(\Phi)$ ($\mathbf{E}_\mathrm{r}(\Phi)$) is the amplitude of the incoming (reflected) pump field for a given polarization angle $\Phi$. The reflected field is computed by decomposing $\mathbf{E}_\mathrm{r}$ into $s$ and $p$ components and using the Fresnel equations. In direct comparison of our calculations with the experimental results, we fix $f_V=0.5$. We note that this value is in good agreement with a study by Neppl \textit{et al.} on dielectric screening on the atomic length scale~\cite{neppl2015direct}.

The local effective field driving the electrons inside the graphene sample is also modified by the screening. To account for the screening of the field inside the graphene, we interpolate the fields by approximating the pump field $\mathbf{E}_\mathrm{pump}$ by
\begin{align}
    \mathbf{E}_\mathrm{pump}(\Phi) = f_F \left(\mathbf{E}_\mathrm{in}(\Phi) + \mathbf{E}_\mathrm{r}(\Phi)  \right) + (1 - f_F) \mathbf{E}_\mathrm{t}(\Phi) \ .
\end{align}
Here, $\mathbf{E}_\mathrm{t}(\Phi)$ is the transmitted field amplitude. We chose $f_F=0.5$, assuming that the effective electric field $\mathbf{E}_\mathrm{pump}$ interpolates between the field outside and inside the material. We note again that this value is in good agreement with the study by Neppl \textit{et al.}~\cite{neppl2015direct}. Finally, we stress that while there is a considerable uncertainty of $E_0$, the momentum-space signatures of the Floquet-Volkov interference discussed in the main text are unaffected over a large parameter range (see extended Fig. 2).

From the electric field amplitudes we also obtain the time-dependent vector potentials $\mathbf{A}_\mathrm{eff}(t)$ and $\mathbf{A}_\mathrm{pump}(t)$, which enter the calculation of the trARPES signals.

\begin{figure}[ht]
    \centering
    \includegraphics[width=\textwidth]{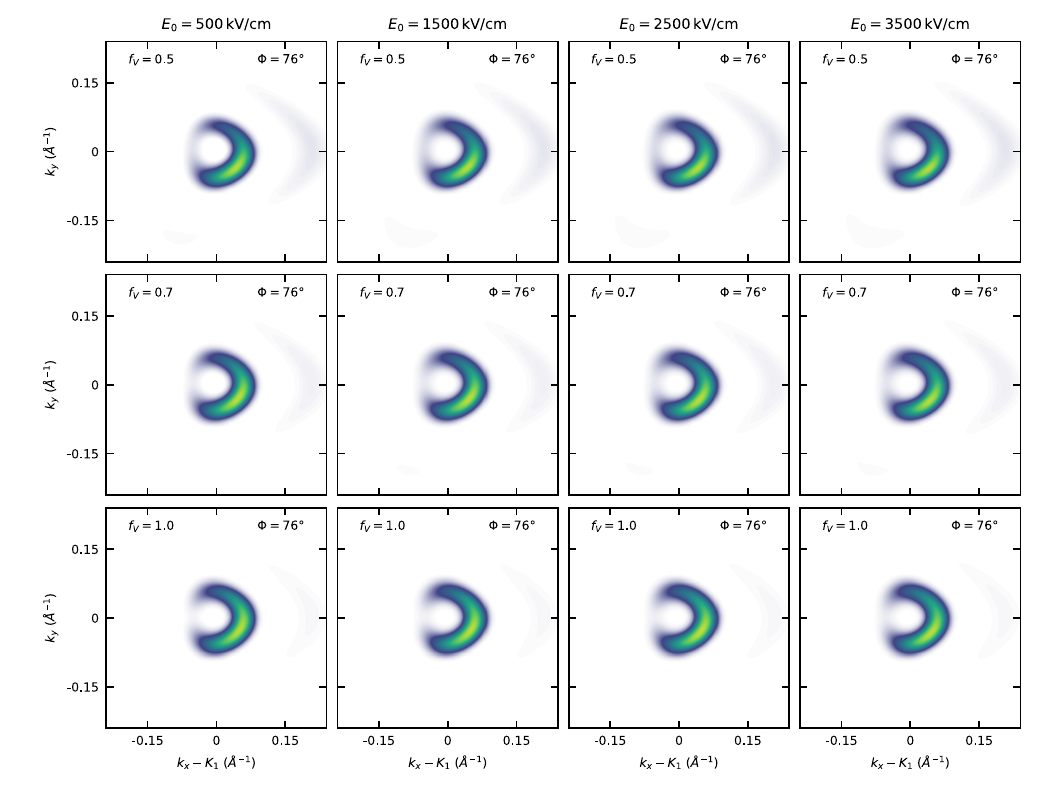} 
    \caption{\textbf{Overview of the calculated Floquet and Volkov $\bm{+1\hbar\Omega}$ sideband momentum fingerprint at $\Phi=76$° as a function of the vacuum electric field strength $\bm{E_0}$ and the factor $\bm{f_V}$.} 
    From top to bottom, the parameter $f_V$ is varied from 0.5 to 1. From left to right, the vacuum electric field strength $E_0$ is varied from 0.5~MV/cm to 3.5~MV/cm. The Floquet scaling factor is set to $f_F=0.5$ for all shown momentum maps. For all figures in the main text, we choose $f_F=f_V = 0.5$ and $E_0 = 3~\mathrm{MV/cm}$.}
    \label{fig:allcones_fit}
\end{figure}

\subsection{Time-dependent dynamics}

The pump-induced dynamics are described by solving the equation of motion for the density matrix:
\begin{align}
    \label{eq:eom_dens}
    \frac{d}{dt}\boldsymbol{\rho}(\mathbf{k},t) = - i [\mathbf{H}(\mathbf{k},t), \boldsymbol{\rho}(\mathbf{k},t)] +D[\boldsymbol{\rho}(\mathbf{k},t)]  \ .
\end{align}
Here, $D[\boldsymbol{\rho}(\mathbf{k},t)]$ denotes a scattering term incorporating pure dephasing dynamics with a decoherence time of $T_2 = 20$~fs. The decoherence of the off-diagonal elements of $\boldsymbol{\rho}(\mathbf{k},t) $ is defined with respect to the instantaneous Hamiltonian $\mathbf{H}(\mathbf{k},t)$, as discussed in ref.~\cite{murakami_doping_2022}. The Hamiltonian is formulated in the velocity gauge, which allows us to consistently compute the time-resolved photoemission signal while retaining gauge invariance~\cite{Schuler2021}. In the velocity gauge,
\begin{align}
    \label{eq:hamiltonian}
    H_{\alpha \alpha^\prime}(\mathbf{k},t) = \varepsilon_\alpha(\mathbf{k})\delta_{\alpha \alpha^\prime} - \mathbf{A}_\mathrm{pump}(t) \cdot \mathbf{v}_{\alpha \alpha^\prime}(\mathbf{k}) + \frac{1}{2}\mathbf{A}_\mathrm{pump}(t)^2 \ .
\end{align}
We have computed the electronic band structure $\varepsilon_\alpha(\mathbf{k})$ and the velocity matrix elements $\mathbf{v}_{\alpha \alpha^\prime}(\mathbf{k}) = \langle \psi_{\mathbf{k}\alpha} | \hat{\mathbf{p}} | \psi_{\mathbf{k}\alpha^\prime} \rangle $ using our in-house all-electron density-functional theory (DFT) code. The consistency with the standard codes \textsc{Quantum Espresso} and \textsc{Wannier90} has been checked. We included the two Dirac-like bands $\alpha=1,2$ in Eq.~\eqref{eq:eom_dens} and \eqref{eq:hamiltonian}.

The pump pulse is parameterized by a Gaussian pulse,
\begin{align}
    \label{eq:avect_pump}
    \mathbf{A}_\mathrm{pump}(t) = \frac{1}{\Omega} S(t)\mathrm{Re}\left[\textbf{}\mathbf{E}_\mathrm{pump} e^{-i\Omega t}\right] \ ,
\end{align}
where $S(t)$ is a Gaussian function with FWHM = 100~fs. 

We have also performed calculations for the Floquet band structure of the Hamiltonian~\eqref{eq:hamiltonian} by replacing $S(t) \rightarrow 1$ and analyzing the thus time-periodic Hamiltonian.

\subsection{Simulation of time-resolved ARPES data}

From the time-dependent density matrix $\boldsymbol{\rho}(\mathbf{k},t)$ we computed the time-resolved photoemission spectra through the time-dependent nonequilibrium Green's function (td-NEGF) formalism. As described in refs.~\cite{Schueler20prx,Schuler2021}, we employed the generalized Kadanoff-Baym ansatz (GKBA) that yields the lesser Green's function from the equation of motion
\begin{align}
    [i \partial_t - \mathbf{H}(\mathbf{k},t)] \mathbf{G}^<(\mathbf{k},t,t^\prime) = 0
\end{align}
with $\mathbf{G}^<(\mathbf{k},t,t) = i \boldsymbol{\rho}(\mathbf{k},t)$. From the Green's function we can then compute the photoemission signal as a function of quasi-momentum $\mathbf{k}$, final-state energy $E$, and pump-probe delay $\tau$ as
\begin{align}
    \label{eq:trarpes}
    I(\mathbf{k},E,\tau) &\propto \mathrm{Im}\sum_{\alpha\alpha^\prime} M^*_{\alpha}(\mathbf{k},E)M_{\alpha^\prime}(\mathbf{k},E)\int^\infty_0 \! dt \int^t_0 \! dt^\prime s(t,\tau) s(t^\prime,\tau) e^{-i \varphi(\mathbf{k},t,t^\prime)} G^<_{\alpha^\prime\alpha}(\mathbf{k},t^\prime, t) \ .
\end{align}
In Eq.~\eqref{eq:trarpes}, $s(t,\tau)$ denotes the envelope function of the probe pulse (taken as Gaussian function with FWHM = 20~fs), while the phase factor is defined by
\begin{align}
    \varphi(\mathbf{k},t,t^\prime) = \int^t_{t^\prime}d \bar{t} \left[\varepsilon_f(\bar{t}) - \omega_\mathrm{pr}\right] \ ,
\end{align}
where $\omega_\mathrm{pr}$ is the photon energy of the probe pulse and $\varepsilon_f(\bar{t})$ is the light-dressed final state energy:
\begin{align}
    \label{eq:final_state_en}
    \varepsilon_f(t) = \frac{\mathbf{p}^2}{2} - \mathbf{A}_\mathrm{eff}(t) \cdot \mathbf{p} + \frac12 \mathbf{A}_\mathrm{eff}(t)^2 \ .
\end{align}
Here, $\mathbf{p}_\parallel = \mathbf{k}$, while $p_\perp$ is determined from $\mathbf{p}^2/2 = E$. In our theory, the LAPE effect can be switched off by replacing $\mathbf{A}_\mathrm{eff} \rightarrow 0$ in Eq.~\eqref{eq:final_state_en}. Similarly, the case of pure Volkov side bands can be simulated by replacing $\mathbf{A}_\mathrm{pump}\rightarrow 0$ in the time-dependent Hamiltonian~\eqref{eq:hamiltonian}.

Our DFT code also allows us to compute the photoemission matrix elements
\begin{align}
    \label{eq:mel}
    M_{\alpha}(\mathbf{k},E) = \langle \chi_{\mathbf{k},E} | \mathbf{e}_\mathrm{pr}\cdot \hat{\mathbf{p}} | \psi_{\mathbf{k}\alpha}\rangle \ ,
\end{align}
where $\mathbf{e}_\mathrm{pr}$ denotes the polarization of the $p$-polarized probe pulse, and where $|\chi_{\mathbf{k},E}\rangle$ are the inverse LEED states. The predictive power of our method to directly compute the matrix elements~\eqref{eq:mel} has been confirmed by comparing the calculations to probe photon-energy dependent measurements in equilibrium.

Combining the light-matter coupling encoded in Eq.~\eqref{eq:hamiltonian} with the matrix elements~\eqref{eq:mel} provides us with an \emph{ab initio} description of time-resolved photoemission~\eqref{eq:trarpes}.

\clearpage


\section{Floquet energy gaps $\Delta E$ in time-resolved ARPES experiments}
\begin{figure}[b]
    \centering
    \includegraphics{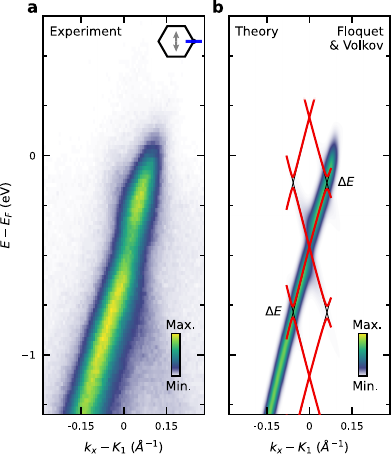}
    \caption{\textbf{Floquet energy gaps $\Delta E$ in ARPES experiments with spectrally broad femtosecond laser pulses.}
    \textbf{a} Energy- and momentum-resolved photoemission spectrum of light-driven graphene ($s$-polarized) collected in the proximity of the K$_1$ point along the $k_x$ momentum-direction (inset). The parameters of the driving light field (the probing light field) are 650~meV, 100~fs, 3~MV/cm (26.5~eV, 20~fs).
    \textbf{b} Floquet energy spectra (red) in the proximity of the K$_1$ point calculated for the experimental conditions predict an energy gap of $\Delta E=70$~meV. As a guide to the eye, the black lines indicate the main band and the $\pm1$st order sidebands neglecting hybridisation. The color-coded data is obtained by calculating an ARPES map for the experimental conditions based on the time-dependent Green's function formalism. The width of the spectral weight dominantly results from the finite width of the ultrashort EUV probe pulse.
    }
    \label{fig:floquetenergygaps}
\end{figure}

In the main text, we highlight two options to observe a Floquet-engineered band structure in an ARPES experiment. Option (ii)  is the identification of light-induced energy gaps where Floquet bands of different photon orders cross ($\Delta E$ in Fig.~1a). For this, extended Fig.~\ref{fig:floquetenergygaps}a shows energy-momentum-resolved photoemission data taken along the k$_x$ momentum direction for $s$-polarized IR pulses (K$_1$ point, $\Phi=90$°), i.e., the momentum direction where Floquet energy gaps are expected~\cite{mahmood_selective_2016,Sentef15natcom,Aeschlimann21nanoletters}. However, the data shows no clear indication for energy gaps.

To verify whether this result is caused by the limited energy resolution in our ultrafast ARPES experiment with spectrally broad laser pulses, we compare the experimental data with our model calculations. For this, we first calculate the energy-momentum dispersion of the Floquet eigenenergies (red lines in extended Fig.~\ref{fig:floquetenergygaps}b). For the 650~meV IR pulses with a vacuum electric field strength of 3~MV/cm, we extract a Floquet energy gap of $\Delta E=70$~meV. Second, we calculate an ARPES spectrum of the light-driven band structure within the non-equilibrium Green's function formalism (color-coded data in extended Fig.~\ref{fig:floquetenergygaps}b). The ARPES spectrum now intrinsically shows a distinct energy broadening, which is mainly caused by the spectrally broad pump and probe laser pulses. We note that the linewidth of the calculated ARPES signature is much narrower than found in our experimental results, indicating that also other broadening effects contribute to the experimental broadening (155~meV, extended Fig.~\ref{fig:crosscorrelation}d), which are not captured in the model. Nevertheless, also in the calculated ARPES spectrum, it is not straightforwardly possible to identify a clear spectroscopic signature of an energy gap. In consequence, for the parameters of the driving light field that are currently accessible with our setup, we conclude that the energy resolution of the time-resolved momentum microscopy experiment is simply not sufficient to directly resolve the spectroscopic signatures of energy gaps.

\clearpage

\bibliography{bibtexfile.bib}

\end{document}